\documentstyle[emulateapj,apjfonts,epsfig]{article}

\newcommand{\sax}{SAX~J1808.4$-$3658}

\begin{document}

\title{A Brown Dwarf Companion for the Accreting Millisecond Pulsar \sax}

\author{Lars~Bildsten}
\affil{\footnotesize Institute for Theoretical Physics and Department of 
   Physics, University of California, \\ Santa Barbara, CA 93106;
   bildsten@itp.ucsb.edu}

\and

\author{Deepto~Chakrabarty}
\affil{\footnotesize Department of Physics and Center for Space Research, 
   Massachusetts Institute of Technology, \\ Cambridge, MA 02139;   
   deepto@space.mit.edu}

\begin{abstract}

  The {\em BeppoSAX} Wide Field Cameras have revealed a population of
faint neutron star X-ray transients in the Galactic bulge.  King
conjectured that these neutron stars are accreting from brown dwarfs
with a time-averaged mass transfer rate $\langle \dot M\rangle \approx
10^{-11} M_\odot$~ yr$^{-1}$ that is low enough for accretion disk
instabilities. We show that the measured orbital parameters of the 401
Hz accreting millisecond pulsar \sax\ support this hypothesis.  A
main-sequence mass donor requires a nearly face-on inclination and a
higher $\langle \dot M \rangle$ than observed, and can thus be
excluded. However, the range of allowed inclinations is substantially
relaxed and the predicted $\langle \dot M \rangle$ is consistent with
that observed if a hot $0.05 M_\odot$ brown dwarf is the donor.

The remaining puzzle is explaining the brown dwarf radius required
(0.13 $R_\odot$) to fill the Roche lobe. Recent observational and
theoretical work has shown that all transiently accreting neutron
stars have a minimum luminosity in quiescence set by the time-averaged
mass transfer rate onto the neutron star.  We show here that the
constant heating of the brown dwarf by this quiescent neutron star
emission appears adequate to maintain the higher entropy implied by a
0.13 $R_\odot$ radius. All of our considerations very strongly bolster
the case that \sax\ is a progenitor to compact millisecond radio
pulsar binaries (e.g. like those found by Camilo and collaborators in
47 Tuc). The very low $\langle \dot M \rangle$ of \sax\ implies that
the progenitors to these radio pulsars are long-lived ($\sim {\rm
Gyr}$) transient systems, rather than short-lived ($\sim {\rm Myr}$)
Eddington-limited accretors. Hence, the accreting progenitor
population to millisecond radio pulsars in 47 Tuc could still be
present and found in quiescence with {\it Chandra}.

\end{abstract}


\bigskip
\centerline{\bf To Appear in The Astrophysical Journal} 
\bigskip

\section{INTRODUCTION}

X-ray monitoring of the Galactic bulge with the {\em BeppoSAX} Wide
Field Cameras (WFCs) has identified seven extremely low-luminosity,
transient X-ray sources (Heise et al. 1998).  All show type I X-ray
bursts from unstable thermonuclear burning, indicating that they are
accreting neutron stars (NSs) with weak ($\ll 10^{10}$ G) surface
magnetic field strengths.  Their peak fluxes (apart from the type I
bursts) are below 100 mCrab, corresponding to a luminosity of $<
5\times 10^{36}$~erg~s$^{-1}$ at the Galactic center.  At least three
recur, with week-long outbursts coming at intervals ranging from
months to years. King (2000) has shown that these sources form a
distinct class of NSs in low-mass X-ray binaries (LMXBs). Their
unusually low time-averaged mass transfer rates ($\langle \dot M
\rangle \sim 10^{-11} M_\odot$ yr$^{-1}$) allow for accretion disk
instabilities to operate and cause the recurrent outbursts. Such low
$\langle \dot M \rangle$'s occur when orbital angular momentum is lost
via gravitational radiation (GR) and when the donor mass is in the
brown dwarf (BD) regime ($\lesssim 0.08M_\odot$; Verbunt \& van den
Heuvel 1995, King 2000). This is the natural endpoint of mass
transfer evolution for a low-mass  donor (e.g. Rappaport,
Verbunt \& Joss 1983). 

  A famous member of this class is \sax, the only known
accretion-powered millisecond pulsar. It was first detected as a 
X-ray transient with type I X-ray bursts in 1996 September by the WFCs
in a 20 day outburst (in~'t Zand et al. 1998). A recent reanalysis of
these data has yielded a distance of 2.5~kpc from the photospheric
radius expansion observed during the type I bursts (in~'t Zand et
al. 2001).  During a second outburst in 1998 April, observations with
the {\em Rossi X-Ray Timing Explorer (RXTE\/)} revealed persistent 401
Hz X-ray pulsations (Wijnands \& van der Klis 1998) and 2~hr binary
motion (Chakrabarty \& Morgan 1998, hereafter CM98).

  The measured orbital parameters of this system allow us to solve for
the binary companion's mass, $M_{\rm c}$, and radius, $R_{\rm c}$, as
a function of orbital inclination. White dwarf companions are ruled
out while low-mass hydrogen main sequence companions are allowed only
for highly improbable inclination values and an $\langle \dot M
\rangle$ inconsistent with the observations (CM98). In this paper, we
show that a more consistent solution involves a $M_{\rm c}\approx
0.05M_\odot$ BD as the mass donor.  This substantially relaxes the
inclination to more likely values and gives an $\langle \dot M
\rangle$ that agrees with observations.  The puzzle is why such a
low-mass donor would fill the Roche lobe at an orbital period of 
2 hours. 

  We conjecture here that the larger entropy implied by the $R_{\rm
c}\approx 0.13R_\odot$ radius can be maintained by the continuous
heating of the BD in quiescence by the thermal emission from the
NS. This naturally makes this system a progenitor to the fast
millisecond radio pulsars (MRPs) in short ($<6$ hr) orbits and shows
that there is a $\sim {\rm Gyr}$ phase of accretion at very low
rates. This is in contrast to progenitor scenarios (e.g. Ruderman,
Shaham and Tavani 1989) which give $\dot M$'s at least two orders of
magnitude higher.

\section{Constraints on the Binary Inclination} 

The binary parameters of \sax\ were precisely measured, with orbital
period $P_{\rm orb}=2.01$~hr and a projected radius $a_{\rm x}\sin i=
62.8$~light-ms (CM98). The mass function is 
\begin{equation}
\label{eq:fx}
    f_{\rm x} \equiv \frac{(M_{\rm c}\sin i)^3}
                          {(M_{\rm x} + M_{\rm c})^2}
              = \frac{4\pi^2(a_{\rm x}\sin i)^3}{GP_{\rm orb}^2} 
              = 3.8\times 10^{-5} M_\odot , 
\end{equation}
where $i$ is the binary inclination (the angle between the line of
sight and the orbital angular momentum vector) and $M_{\rm x}$ is the
NS mass. Given $M_{\rm x}$
and $i$, equation (\ref{eq:fx}) yields $M_{\rm c}$, as shown in
Figure~1 for $M_{\rm x}=1.4 M_\odot$ and $M_{\rm x}=2 M_\odot$. The
minimum companion mass ($i=90^\circ$) is then $\simeq$0.043 (0.054)
$M_\odot$ for $M_{\rm x}=$1.4 (2.0) $M_\odot$, far below the hydrogen
burning limit of $\approx 0.08 M_\odot$.  For randomly selected
binaries, the a priori probability of observing a system with $i<i_0$
is $(1-\cos i_0)$, so the likeliest solutions are those with
BD donors.

Mass transfer from the Roche-lobe--filling\footnote{Most LMXBs accrete
via Roche lobe overflow. Substantial mass loss through an intrinsic
stellar wind is unlikely for very low-mass donors.} low mass companion
is driven by angular momentum loss from GR, $\dot J_{\rm GR}$ (see
review by Verbunt \& van den Heuvel 1995).  The mass transfer rate is
then $\dot M_{\rm GR}=3M_{\rm c}\dot J_{\rm GR}/2J$, or
\begin{equation} 
   \dot M_{\rm GR}\approx  3.8\times 10^{-11}\mbox{\rm\ $M_\odot$ yr$^{-1}$\ }
     \left(\frac{M_{\rm c}}{0.1 M_\odot}\right)^2
     \left(\frac{M_{\rm x}}{1.4 M_\odot}\right)^{2/3}
     \left(\frac{P_{\rm orb}}{\mbox{\rm 2 hr}}\right)^{-8/3} , 
\end{equation}
where we have set\footnote{Equation (2) agrees with King, Kolb and
Szuszkiewicz (1997) but is a factor of three larger than equation (3)
of King (2000). Though we are not certain, it appears that King's
number is lower because he presumed $n=1$ for his mass-radius relation
and dropped a significant figure. Our choice of $n=-1/3$ is for the
isentropic and heated BD companion motivated in \S's 3 and 4.}
$n=-1/3$ in the donor's mass-radius relation, $R_{\rm c}\propto M_{\rm
c}^n$.  This is plotted in the bottom panel of Figure~1.  The mass
transfer rate in equation (2) is smaller by a factor of two for a main
sequence mass-radius relation ($n=1$).

We now compare this to the measured $\langle \dot M \rangle$. 
The 1998 April outburst of \sax\ was monitored by a series of
pointed {\em RXTE} observations, from which a total fluence of
$\approx 4.3\times 10^{-3}$~erg~cm$^{-2}$ (3--150 keV; Gilfanov et
al. 1998) was measured.  The recurrence time of the outbursts is
roughly 20 months (in 't Zand et al. 1998; Marshall 1998; van der
Klis et al. 2000).  Presuming that the 1998 April outburst was
typical, the time-averaged flux is $9\times 10^{-11}$
erg~cm$^{-2}$~s$^{-1}$, giving
\begin{equation}
   \langle \dot M \rangle  \approx 5\times 10^{-12} \mbox{\rm\ $M_\odot$ yr$^{-1}$\ }
      \left(\frac{d}{\mbox{\rm 2.5 kpc}}\right)^2
      \left(\frac{M_{\rm x}}{1.4 M_\odot}\right)^{-1}, 
\end{equation}
which is most consistent for $M_c\approx 0.05M_\odot$, thus implying 
that the binary is not viewed face-on. 
  
 Three other pieces of evidence also suggest that \sax\ is not viewed
face-on.  First, there is a 2\% modulation in X-ray intensity at the
orbital period, with the minimum occurring when the NS is behind the
companion (CM98; Lee, Chakrabarty, \& Morgan 2001, in preparation).
This suggests that $i$ must be large enough to allow partial X-ray
blockage from some circumbinary material. Second, the 2 hour
single-peaked modulation in the optical intensity (Giles, Hill, \&
Greenhill 1999) has been confirmed by Homer et al. (2001).  Flux
minimum occurs when the NS is behind the companion, suggesting that
X-ray heating of the companion is the origin of the modulation; again
implying inclinations other than face-on. Finally, detailed modeling
of the optical companion's multiband photometry during outburst with a
simple X-ray heated disk model suggests that $\cos i<0.45$ (Wang et
al. 2001, in preparation). Though clearly not a face-on system, the
absence of deep X-ray eclipses indicates that we must be viewing \sax\
with $\cos i > 0.15$ (CM98).

  In summary, a consideration of the mass transfer rates and
inclination constraints lead us to conclude that the likely Roche lobe
filling donor has a mass $M_{\rm c}\approx 0.05 M_\odot$.  We now
discuss how such a system is formed and maintains Roche-Lobe
filling at the two hour  orbital period. 

\section{Companion Radius and Binary Evolution} 

  The mean density of a Roche-lobe--filling companion (for the case
where $M_{\rm c}\ll M_{\rm x}$) is set by the binary period (Faulkner,
Flannery, \& Warner 1972), so that $R_{\rm c} = 0.17 ({M_{\rm c}}/0.1
{M_\odot})^{1/3}\ R_\odot$, as shown by the dark solid curve in
Figure~2.  Also plotted are the mass-radius relations for cold helium
white dwarfs (Paczynski 1967), low-mass hydrogen main sequence stars
(Tout et al. 1996), and BDs of ages 0.1, 0.5, 1.0 and 5.0 Gyr
(Chabrier et al. 2000).  Note that the mass-radius relation for any
Roche-lobe--filling solution allowed by the observations must
intersect (or at least lie near) the dark solid curve in Figure~2.
Thus, we see immediately that cold helium white dwarfs are too small
(for any mass).  Low-mass hydrogen main sequence stars give a solution
at $M_c=0.17 M_\odot$, corresponding to $\cos i =0.95$.  This solution
has a small {\it a priori} probability (5\%); moreover, it would give
a mass transfer rate (from equation (2) with $n=1$) of $\langle \dot M
\rangle> 5\times 10^{-11} M_\odot \ {\rm yr^{-1}}$, a factor of ten higher
than observed.  Thus, the solution we consider most likely is a
$0.05M_\odot$ BD with radius $R_c\approx 0.13R_\odot$.

  Many (e.g. Paczynski \& Sienkiewicz 1981; Rappaport, Verbunt \& Joss
1983) have shown that the mass transfer evolution (under GR losses) of
a $M<M_\odot$ main sequence donor leads to a BD companion within a few
Gyrs. The initial evolution of these systems is towards shorter
orbital periods when the Kelvin-Helmholtz (KH) time is shorter than
the mass transfer time scale (this is typically when the donor is on
the main sequence).  Once the KH time becomes longer than the mass
transfer time scale, the donor expands under further mass loss (if
adiabatic, then $n=-1/3$), and the period increases.  This sets the
minimum orbital period at $\approx 80$ minutes and if the BD is
allowed to cool once nuclear burning has ceased, the companion would
have a mass $\ll 0.05 M_\odot$ when a period of 2 hours is reached
(e.g. Rappaport et al. 1983).  Hence, some inhibition of the
contraction must occur in \sax.

 The issue of expanding BD companions by external heating (or driving
winds) came to a sharp focus with the discovery of the eclipsing
optical counterpart to PSR 1957+20 (Fruchter et al. 1990). Many models
were constructed, some of which relied on tidal heating of a low-mass,
asynchronous object for the expansion (e.g. Applegate and Shaham
1994). Rasio, Pfahl, and Rappaport (2000) present a similar scenario
for the MRPs in 47 Tuc in short orbital periods, implying mass
transfer rates a factor of 100-1000 higher than observed in \sax.
Ruderman et al. (1989) had previously considered the possibility that
the light companion was heated at a rate adequate to drive high mass
(and angular momentum) loss that would increase the mass transfer rate
in the binary to near-Eddington limits (see Bhattacharya 1995 for  a
review). Some of their motivation was
the lack of observed low-accretion rate LMXB's. The {\em BeppoSAX}
discoveries have clearly changed the situation and motivated us to
reconsider GR-driven systems at late times when the donor mass is
below the hydrogen burning limit. 

\section{The Thermostat of X-Ray Heating in Quiescence} 

The companion is heated to $>2\times 10^4$~K during the outbursts
in \sax\ .  We presume that this transient external heating has 
little effect
on the interior of the BD; rather we focus on the continuous heating
in quiescence from the hot NS. Bildsten and Brown (1997) showed that
the quiescent heating of the outer disk was insufficient to suppress
the disk instability responsible for the outbursts. However, it
appears sufficient to dramatically alter the entropy evolution of the
low-mass stellar companion once nuclear burning has ceased. 

 Transiently accreting NSs are always detected in quiescence at an
X-ray luminosity $>10^{32}$~erg~s$^{-1}$ (see Bildsten and Rutledge
2000 for an overview). This emission has two components: a soft
thermal component from the NS surface (Brown, Bildsten \& Rutledge
1998; Rutledge et al. 1999, 2000, 2001) and a hard power-law tail of
unknown origin (see Campana \& Stella 2000 for a discussion). The
thermal emission level is well predicted by the relation of Brown et
al. (1998)
\begin{equation}
\label{eq:lq}
L_{\rm th}\approx 6\times 10^{32} {\rm erg \
s^{-1}}\left(\langle\dot M\rangle\over 10^{-11} 
M_\odot \ {\rm yr^{-1}}\right), 
\end{equation} 
which arises from the deep nuclear heating that
the NS receives every outburst. 

 For the time-averaged flux derived above, equation (\ref{eq:lq})
predicts thermal emission at $4.5\times 10^{-13} {\rm erg \ cm^{-2} \
s^{-1}}$. There are two weak detections of the quiescent X-ray
emission from \sax.  Stella et al. (2000) report $(1-2)\times 10^{-13}
{\rm erg \ cm^{-2} \ s^{-1}}$ (0.5--10 keV) with {\em BeppoSAX} (in
excellent agreement once the spectral corrections are applied; see
Table 4 of Rutledge et al. 2000) while Dotani, Asai \& Wijnands (2000)
report a value about a factor of 3 lower from {\em ASCA}.

We use the quiescent luminosity from equation (\ref{eq:lq}) as a
lower-limit to the heating and find the temperature, $T_{\rm H}$, of
the companion on the side facing the NS, $\sigma_{SB}T_{\rm
H}^4=L_{\rm th}/4\pi(a-R)^2$, as a function of $\langle \dot
M\rangle$.  For plausible inclinations of \sax\ , $T_{\rm H}\approx
3800-4800 {\rm K}$, far above the expected value for a BD that has
cooled after passing through the period minimum. We conjecture that
this continuous heating fixes the entropy of the BD at a value higher
than would occur in the absence of heating, where the entropy is free
to decrease once nuclear burning has halted. This would also increase
the minimum orbital period of such binaries.

There are no appropriate calculations that allow us to accurately
estimate the slowing of contraction from partial irradiation at these
levels.  However, the effects of less irradiation of smaller mass
objects, the irradiated extrasolar giant planets, lead to radii 50\%
larger than expected (Burrows et al. 2000), as has been observed for
the transiting planet around HD 209458 (Brown et al. 2001).

\section{Conclusions and Comparisons to Millisecond Pulsars} 

  In addition to making the case that the companion to \sax\ is a
brown dwarf, our work also places it in relation to the short orbital
period ($<6 \ {\rm hr}$) fast ($P_s< 5 \ {\rm ms}$) MRPs 
both in the field (PSR J2051$-$0827, Stappers et. al. 1999)
and in globular clusters (47~Tuc I, J, O, P and R, Camilo et
al. 2000; PSR J1807$-$2459 in NGC~6544, Ransom et. al. 2001;  PSR
1908+00 in NGC~6760, Deich et al. 1993). The longer orbital period for
PSR 1957+20 places it far outside of this group. The top panel in
Figure~3 shows $M_c$ for $\cos i=0.5$ for the field pulsar PSR
J2051$-$0827 (open triangle) and globular cluster pulsars (solid
squares).  All have lower mass companions than \sax\ (asterisk).

  We plot two other related quantities in Figure 3. We use the
measured orbital parameters and presume that the companion was, in the
past, Roche lobe filling at the current orbital period and
transferring mass at the rate set by GR losses with $n=-1/3$. The mass
transfer rate (shown as a time scale in the bottom panel) and the
heated companion temperature, $T_{\rm H}$ (middle panel) are shown for
$\cos i=0.5$.  \sax\ has the most heated companion and nearly the
shortest accretion time scale (47 Tuc R is just barely shorter),
consistent with it being a progenitor to the pulsars. 

 If the heating maintains a constant entropy in the BD of \sax\ ,
further mass transfer will evolve it along the solid line. Since the
heating is clearly dropping, we also plot (dashed line) a constant
radius evolution ($n=0$, or decreasing entropy). These clearly bracket
many of the observed systems at longer orbital periods.  For $n=0$, it
will take 6.6 Gyrs for \sax\  to reach a system like PSR J2051-0827
($M_c=0.03M_\odot$).

Something eventually allows the BD to contract and fall within the
Roche lobe, halting mass transfer and allowing the NS to become a
MRP. In this respect, it is important that most of the $T_{\rm H}$'s
for the MRPs are comparable to a BD in the contraction phase (Chabrier
et al. 2000), implying that the reduced heating at larger orbital
periods might well allow the BD to finally fall within the Roche
radius.

  \sax\ is clearly a long-lived ($\sim {\rm Gyr}$) X-ray source that
undergoes mass transfer at a rate of $\approx 10^{-11}
M_\odot$~yr$^{-1}$.  If this long-lived system is the progenitor type
to the short--orbital-period MRPs in 47~Tuc, then we expect to see
faint X-ray transients in 47~Tuc comparable in number to the MRP
binaries. The infrequent sampling and long recurrence times would
make their detection in outburst unlikely. However, they should easily
be detected in quiescence with {\it Chandra} at the levels implied by
equation (\ref{eq:lq}).  Discovery of such a plethora of quiescent, 
short--orbital-period LMXBs would confirm this connection.

\acknowledgments

  We thank D. Chernoff, A. Cumming, S. Phinney, R. Rutledge, H. Spruit,
S. Thorsett, and M. van
Kerkwijk for conversations and comments on this work, which was 
partially supported by NASA via grants NAG 5-8658 and NAG 5-9184 and
by the NSF under grant PHY99-07949. L. B. is a
Cottrell Scholar of the Research Corporation.

\vfill\eject

\begin{figure}
\plotone{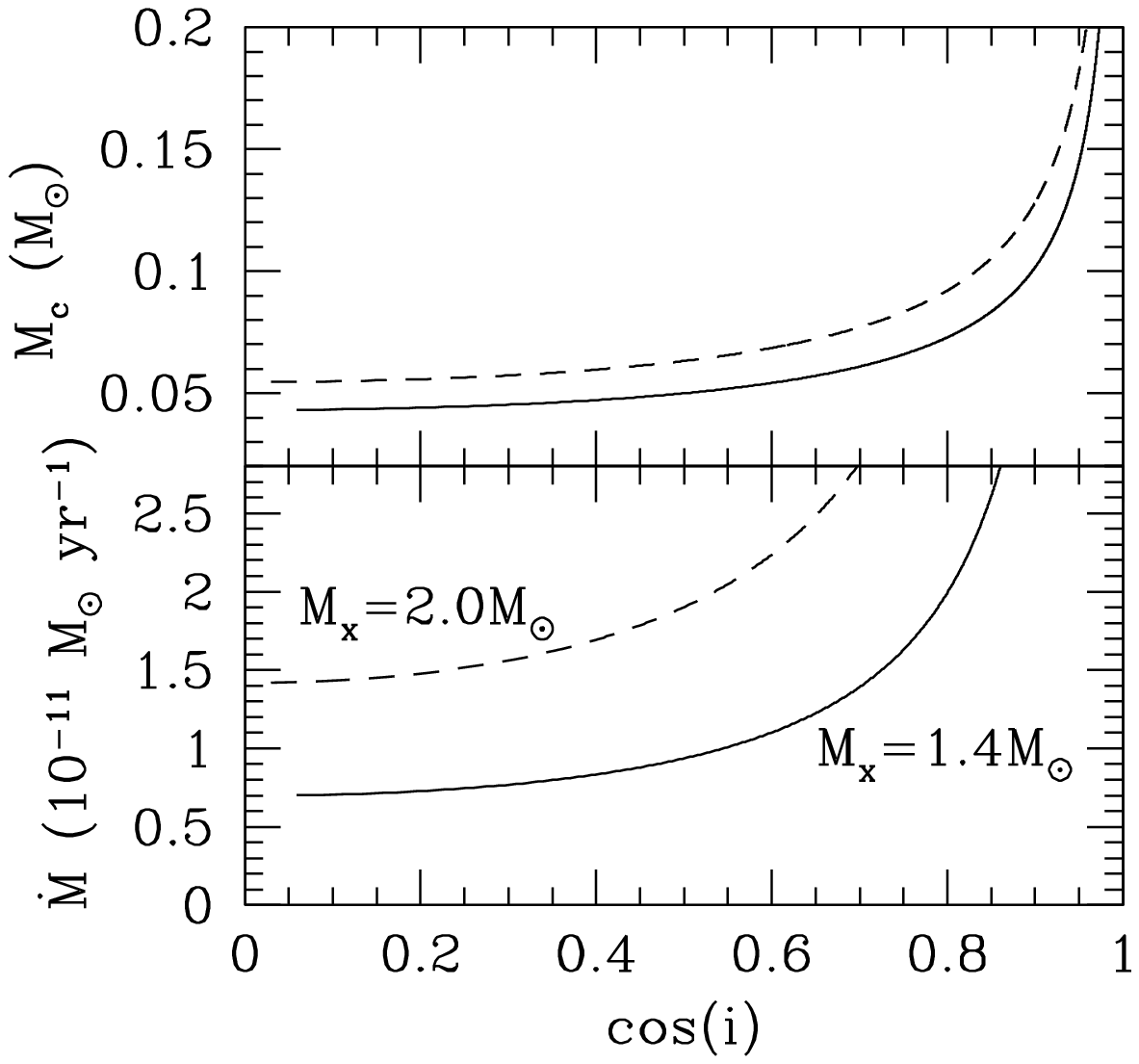}
\figcaption{The companion mass and mass transfer rate 
for \sax\ as a function of inclination angle. The solid (dashed) 
lines are for a neutron star mass of $1.4(2.0)M_\odot$. 
\label{fig:mass}}
\end{figure}

\begin{figure}
\plotone{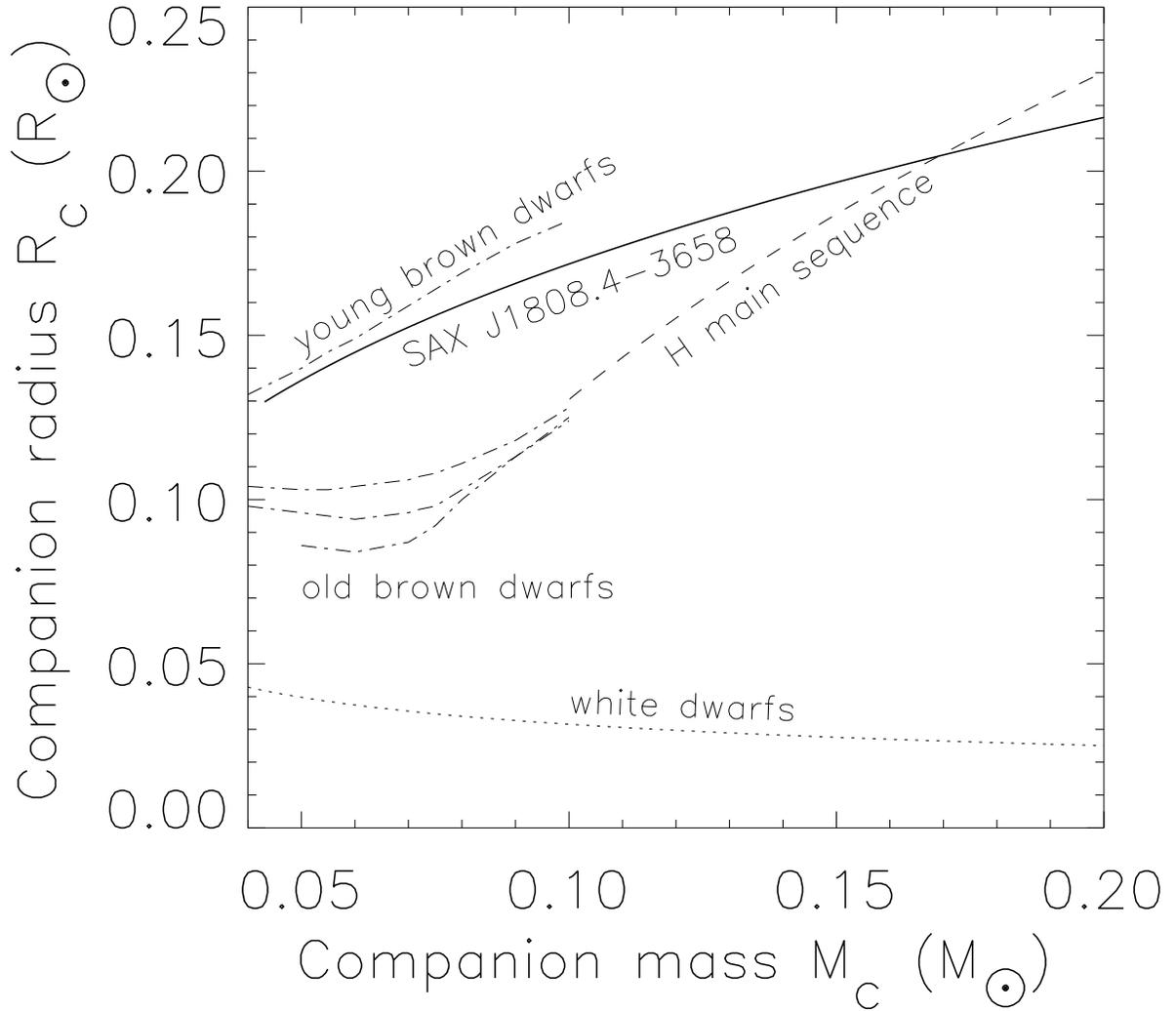}
\figcaption{
Possible companion types for \sax.  The dark solid curve shows the
mass-radius relation for \sax\ allowed by the observations.  The other
curves show the theoretical mass-radius relations for cold He white
dwarfs ({\em dotted curve}), low-mass hydrogen mass sequence stars
({\em dashed curve}), and brown dwarfs of ages ({\em dot-dash
curves}) of 0.1, 0.5, 1.0 and 5.0 Gyr, from top to bottom.
\label{fig:massr}}
\end{figure}

\begin{figure}
\plotone{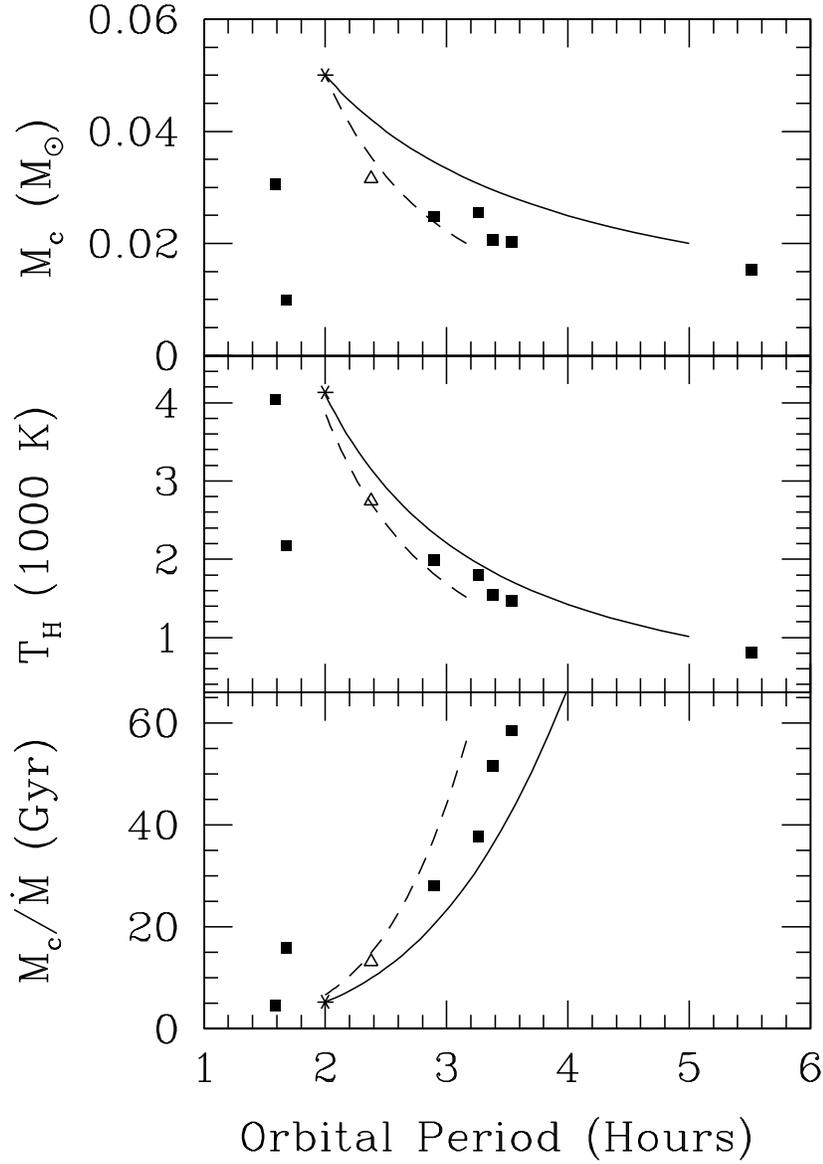}
\figcaption{Companion masses, heated companion temperature and mass
transfer rates for \sax\ ({\em asterisk}) and the short orbital period
millisecond radio pulsars in the field (PSR J2051-0827, {\em open triangle}) and in
globular clusters ({\em filled squares}), all presuming a NS mass of
$1.4M_\odot$. All points use the measured orbital parameters and
assume that $\cos i=0.5$. The solid (dashed) line denotes the future trajectory for \sax\ 
when  $n=-1/3$ (n=0). 
\label{fig:ver3}}
\end{figure}

\end{document}